\documentclass[8.5pt,twoside,twocolumn]{article}
\usepackage[super,sort&compress,comma]{natbib} 
\usepackage{mhchem}
\usepackage{times}
\usepackage{amsmath}
\usepackage{sectsty}
\usepackage{balance} 
\usepackage{bm}
\usepackage{color}
\usepackage{graphicx} %eps figures can be used instead
\usepackage[format=plain,justification=default,singlelinecheck=off,font=small,labelfont=bf,labelsep=space]{caption} 
\usepackage{fancyhdr}
\usepackage{setspace}

\begin{document}

\thispagestyle{plain}
\fancypagestyle{plain}{
\fancyhead[L]{\includegraphics[height=8pt]{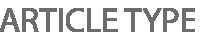}}
\fancyhead[C]{\hspace{-1cm}\includegraphics[height=20pt]{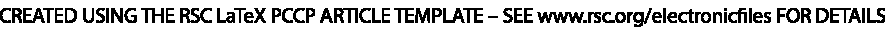}}
\fancyhead[R]{\includegraphics[height=10pt]{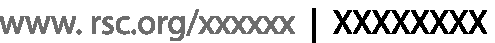}\vspace{-0.2cm}}
\renewcommand{\headrulewidth}{1pt}}
\renewcommand{\thefootnote}{\fnsymbol{footnote}}
\renewcommand\footnoterule{\vspace*{1pt}% 
\hrule width 3.4in height 0.4pt \vspace*{5pt}} 
\setcounter{secnumdepth}{5}

\makeatletter 
\def\subsubsection{\@startsection{subsubsection}{3}{10pt}{-1.25ex plus -1ex minus -.1ex}{0ex plus 0ex}{\normalsize\bf}} 
\def\paragraph{\@startsection{paragraph}{4}{10pt}{-1.25ex plus -1ex minus -.1ex}{0ex plus 0ex}{\normalsize\textit}} 
\renewcommand\@biblabel[1]{#1}            
\renewcommand\@makefntext[1]% 
{\noindent\makebox[0pt][r]{\@thefnmark\,}#1}
\makeatother 
\renewcommand{\figurename}{\small{Fig.}~}
\sectionfont{\large}
\subsectionfont{\normalsize} 

\fancyfoot{}
\fancyfoot[LO,RE]{\vspace{-7pt}\includegraphics[height=9pt]{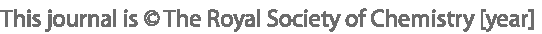}}
\fancyfoot[CO]{\vspace{-7.2pt}\hspace{12.2cm}\includegraphics{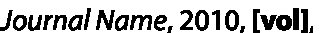}}
\fancyfoot[CE]{\vspace{-7.5pt}\hspace{-13.5cm}\includegraphics{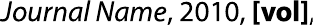}}
\fancyfoot[RO]{\footnotesize{\sffamily{1--\pageref{LastPage} ~\textbar  \hspace{2pt}\thepage}}}
\fancyfoot[LE]{\footnotesize{\sffamily{\thepage~\textbar\hspace{3.45cm} 1--\pageref{LastPage}}}}
\fancyhead{}
\renewcommand{\headrulewidth}{1pt} 
\renewcommand{\footrulewidth}{1pt}
\setlength{\arrayrulewidth}{1pt}
\setlength{\columnsep}{6.5mm}
\setlength\bibsep{1pt}

\twocolumn[
  \begin{@twocolumnfalse}
    \noindent\LARGE{\textbf{Generic model for tunable colloidal aggregation in multidirectional fields}}
    \vspace{0.6cm}

    \noindent\large{\textbf{Florian Kogler,$^{\ast}$\textit{$^{a}$} Orlin. D. Velev,\textit{$^{b}$} Carol K. Hall,\textit{$^{b}$} and
    Sabine H. L. Klapp\textit{$^{a}$}}}\vspace{0.5cm}
    %Please note that \ast indicates the corresponding author(s) but no footnote text is required. 
    %\footnote{test}

    \noindent\textit{\small{\textbf{Received Xth XXXXXXXXXX 20XX, Accepted Xth XXXXXXXXX 20XX\newline
    First published on the web Xth XXXXXXXXXX 200X}}}

    \noindent \textbf{\small{DOI: 10.1039/b000000x}}
    \vspace{0.6cm}
    %Please do not change this text.

    \noindent \normalsize{Based on Brownian Dynamics computer simulations in two dimensions
    we investigate aggregation scenarios of colloidal particles
    with directional interactions induced by multiple external fields. To this end we propose a model
    which allows continuous change in the particle interactions from point-dipole-like to patchy-like
    (with four patches). We show that, as a result of this change, the non-equilibrium aggregation
    occurring at low densities and temperatures transforms from conventional diffusion-limited cluster aggregation
    (DLCA) to slippery DLCA involving rotating bonds;
    this is accompanied by a pronounced change of the underlying
    lattice structure of the aggregates from square-like to hexagonal ordering. Increasing the temperature we find
    a transformation to a fluid phase, consistent with results of a simple mean-field density functional theory.}
    \vspace{0.5cm}
 \end{@twocolumnfalse}
 ]

\section{Introduction}

\footnotetext{\textit{$^{a}$~Institute of Theoretical Physics, Technical University of Berlin, Hardenbergstr. 36, 10623 Berlin, Germany. E-mail: kogler@physik.tu-berlin.de}}
\footnotetext{\textit{$^{b}$~Department of Chemical and Biomolecular Engineering, North Carolina State University, Raleigh, North Carolina, 27695-7905, USA. }}

Recent progress in the synthesis and directional binding of nanometer to micrometer sized patchy and anisotropic particles
makes possible the assembly of colloidal structures with multiple directed bonds~\cite{Walther2008-janus,Pine2012_directionalcolloids,Salomon2011-review}. 
The directional bonding can also be achieved by permanently embedded or 
field-induced dipole and/or multipole moments allowing directional and selective particle 
bonding~\cite{Velev2004-fielddriectedassembly,Pine2012-shiftdip, Granick2012-shiftdip,rosenthal,Nu2013-molecules}.
Within this class, particles with 
field-induced dipolar interactions~\cite{Velev2004-fielddriectedassembly,gangwal:passive,gangwal:nets,kretzschmar2013-ironcap,orlinBhuvnesh}
are especially interesting because 
switching the fields on and off is equivalent to switching the particle interactions on and off. 
This means that aggregation mechanisms~\cite{loewen2008-fieldreview,loewen2014-magnetic} can be 'dialed in'. 
Furthermore, the orientation of inductive fields may be used to direct particle 
aggregation~\cite{kretzschmar2013-ironcap,orlinBhuvnesh,gangwal:passive,gangwal:nets,Zahn2005-dialedinstructureformation,chandra2014-directedDLA}. 
In consequence, such directed self-assembly processes may be 
exploited for the formation of new functional materials with specific and/or adjustable properties. Hence, 
understanding the interplay between externally induced particle 
properties, external fields and thermodynamic conditions, e.g., temperature, is of fundamental interest in modern material science, 
but also from a statistical physics point of view.

An important subset of the many classes of self-assembled structures are percolated colloidal networks, 
which are characterized by system-spanning cross-linked (patchy) particle clusters that are realizable even 
at low volume fractions~\cite{gangwal:nets,Schmidle2013, Sciortino2010-networkfluid,Schmidle2012-dipolegel}. 
Such network-like aggregates are considered to be the underlying micro-structures of gels and have been 
intensively investigated in experiment and theory under equilibrium as well as non-equilibrium 
conditions~\cite{Zaccarelli-Gelsreview,Sciortino2008-irreversible}. In the latter, qualitatively different aggregation 
mechanisms can be identified, namely diffusion limited cluster aggregation (DLCA)~\cite{Witten1981-DLA,Meakin1995-DLAreview} 
and reaction limited cluster aggregation (RLCA)~\cite{Meakin1988-RLCA2D}. 
In the DLCA regime each particle collision leads to the formation of a rigid and essentially 
(on the timescale of the experiment) unbreakable bond with fixed spatial orientation. 
In contrast, in the RLCA regime the probability to form a rigid bond at collision is small. 
Systems with DLCA undergo irreversible dynamics and form fractal aggregates with specific 
fractal dimensions $D_f \approx 1.71$ in continuous two-dimensional space~\cite{Meakin1989-DLA2D,Meakin1995-DLAreview}.
Such colloidal systems are considered to be 'chemical gels' and can be realized by having particle 
interactions that are much stronger than $k_BT$, preventing particles from dissociating due to thermal 
fluctuations. This leads to a pronounced hindrance of structural reconfiguration of large particle aggregates~\cite{Zaccarelli-Gelsreview,cruz2013-frozenionic}.
However, at higher temperatures these systems become 'physical gels' where single particles and 
larger substructures start to connect and disconnect frequently. This strongly affects (increases) 
the fractal dimension~\cite{Sciortino2008-irreversible,Pusey1995-tempDLA1} and finally allows the system to achieve its equilibrium state.

A recently introduced new type of DLCA, which accounts for local rearrangements via flexible bonds, 
is slippery diffusion limited cluster aggregation (sDLCA)~\cite{Nicolai2008-slippery,Mason2007-slippery}.
Slippery bonds allow particles to move or rotate around each other as long as they stay in contact, 
meaning that bonds are still unbreakable but can change their orientation. 
This additional degree of freedom generates, at least in three-dimensional simulations~\cite{Nicolai2008-slippery,Mason2007-slippery}, 
aggregates of the same fractal dimension as classical 
DLCA but with a larger coordination number.

DLCA processes have been studied extensively in systems with isotropically attractive 
particles~\cite{Witten1981-DLA,Meakin1989-DLA2D} but also in systems with patchy particles bearing 
permanent and/or locally restricted interaction sites on their 
surfaces~\cite{Helgesen1988-dipoleDLA,Ito2005-attractionLCA,Ito2009-dipoleDLAtemp,Biswal2013-magneticfractal}. 
In the latter, the spatial orientations of interaction sites can either be free to 
rotate~\cite{Helgesen1988-dipoleDLA,Sciortino2008-irreversible,Ito2009-dipoleDLAtemp} or fixed in 
space~\cite{Popescu2004-anisotropicDLA,Ball1986-anisotropicDLA,Biswal2013-magneticfractal,Schmidle2013}. 
When the orientations of interaction sites are fixed in space, the associated 'chemical gels' undergo {\em anisotropic} 
diffusion limited aggregation which yields a fractal dimension 
of $D_f \approx 1.5$~\cite{Lee1996-tempDLA2,Popescu2004-anisotropicDLA,Ball1986-anisotropicDLA, Biswal2013-magneticfractal,Ito2009-dipoleDLAtemp}, lower than for the isotropic case. 
This situation occurs, e.g., due to the presence of external 
fields~\cite{Biswal2013-magneticfractal,chandra2014-directedDLA} or 
in lattice models~\cite{Lee1996-tempDLA2,Popescu2004-anisotropicDLA,Ball1986-anisotropicDLA}, 
where motion is naturally restricted to certain directions. 

In the present paper we are particulary interested in the aggregation of colloids with field-induced multipolar interactions.
Examples are capped (metal-coated dielectric particles studied earlier by some of us~\cite{gangwal:nets,Schmidle2013}),
where time-dependent electric fields can induce quadrupolar-like interactions.
Here we consider even more complex interactions caused by {\em crossed} (orthogonal) fields.
We briefly mention two examples of possible experimental realizations of such systems.
The first one is a quasi two-dimensional system of suspended colloidal particles, each composed 
of super-paramagnetic iron-oxide aggregates embedded in a polymer matrix, which has been investigated 
experimentally by one of us \cite{orlinBhuvnesh,Bhuvnesh_nets}. In this case crossed external electric and magnetic fields, 
oriented in plane but perpendicular to each other, can be used to induce independent electric and magnetic 
dipole moments in the colloids leading to a directed self-assembly process resulting in two-dimensional 
single-particle chain networks. 
A second possible experimental and quasi two-dimensional system consists of suspended colloidal particles 
under the influence of two in-plane orthogonal AC electric fields with a phase shift of $\pi$. 
The fields will polarize the particles' ionic layer periodically but at different times due to their phase shift. 
By adjusting the field frequencies and phases to the relevant timescales governing particle diffusion 
and the relaxational dynamics of the polarized ionic layer, two decoupled orthogonal 
dipole moments in each particle can in principle be generated by this setup. 
In both cases, the crossed dipole moments 
might be characterized as point-like or having a finite distance between their 
constitutive charges (or microscopic dipole moment distributions in the magnetic case).

\begin{figure}
\centering
\includegraphics[width=0.45\textwidth]{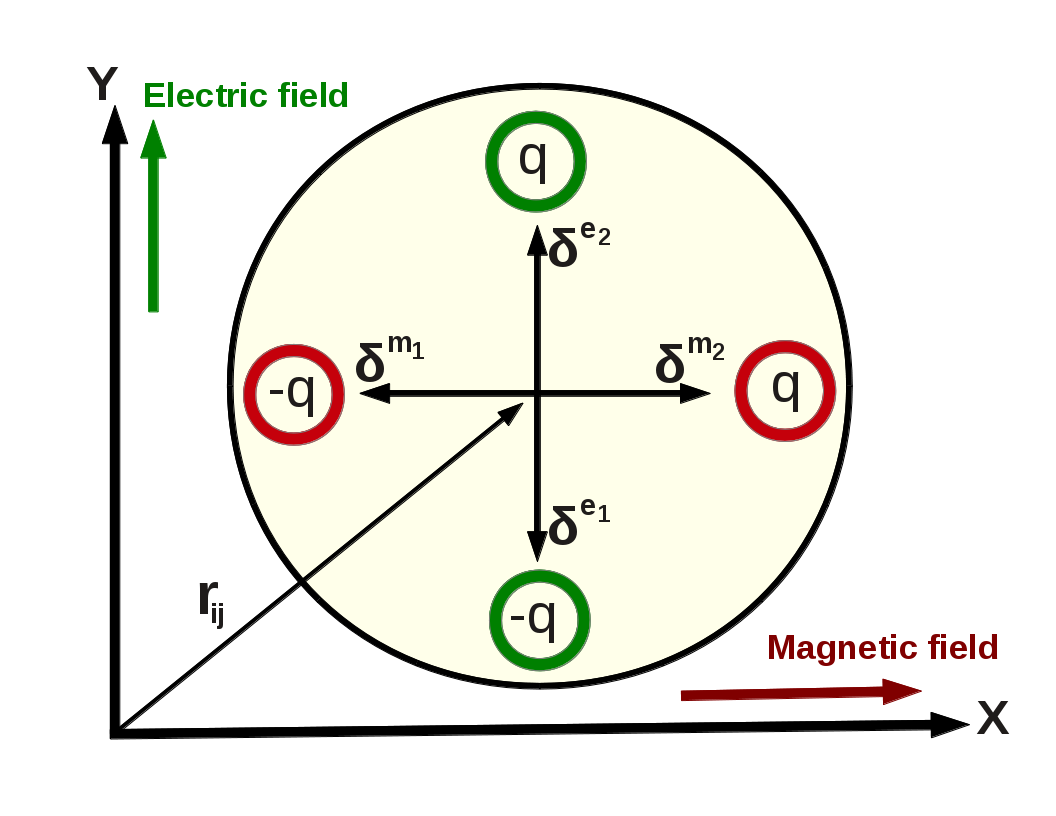}
\caption{(Color online) Distribution of externally induced fictitious "charges" $q$ inside a particle.
Positions of charges are determined by the vectors 
$\bm{\delta}^{\alpha_k} \in [-\delta\mathbf{e}_x,\delta\mathbf{e}_x,-\delta\mathbf{e}_y,\delta\mathbf{e}_y$]) 
pointing either parallel or anti-parallel to the corresponding fields.}
\label{fig:1} 
\end{figure}

Here, we investigate the structure formation in such systems in a conceptional fashion by means of two-dimensional 
Brownian dynamics (BD) simulations of a generic particle model. The idea is to mimick externally-induced 
dipole moments via two pairs of screened Coulomb potentials that are decoupled to account either for magnetic 
and electric interactions or for two temporarily present electric interactions. The two charges associated with each 
pair are shifted outward from the particle center, one parallel to the corresponding field and the other one anti-parallel. 
A sketch of such a particle with its internal arrangement of interaction centers is shown in Fig.~\ref{fig:1}. 
By changing the charge separation, we systematically investigate the (transient) structural ordering and aggregation behavior predicted by this model.

Highlights of our results are the following: At very high interaction energies and large charge separations 
we find that the particles undergo anisotropic diffusion limited cluster aggregation with rectangular local particle arrangements.  
Lowering the charge separation shifts the model behavior to a slippery diffusion limited aggregation (sDLCA) 
regime accompanied by a sharp transition of the lattice structure from rectangular to hexagonal. 
In the proximity of this transition we observe long-lived or arrested frustrated structures consisting of 
strongly interconnected hexagonal and rectangular lattice domains connected with each other. 
We also show that, upon increase of the temperature, the systems enter a fluid state. 
The corresponding 'fluidization' temperature turns out to be very close to the spinodal temperatures
obtained from a mean-field density functional theory.

The rest of this paper is oranized as follows. In section~\ref{Model} we present our model and discuss relevant target quantities
obtained in the BD simulations.
Numerical results are described in section~\ref{Results}, where we discuss first a specific low-temperature, low-density, state and
then turn to the role of temperature and density. Finally, our conclusions are summarized in section~\ref{Conclusions}.

\section{Theoretical Model}
\label{Model} 
We consider a two-dimensional system of $N$ soft spheres of equal diameter $\sigma$. 
The soft sphere interactions are repulsive and are modeled by a shifted and truncated (12,6) Lennard-Jones Potential
\begin{equation}
U_{SS}(\mathbf{r}_{ij}) = 4\epsilon\left((\sigma/ r_{ij})^{12} - (\sigma/ r_{ij})^{6}+1/4\right)
\end{equation}
which is cut off at $r^{c,SS}_{ij} = 2^{1/6}\sigma$. 
Here, $r_{ij}=|\mathbf r_{j}-\mathbf r_{i}|$ is the particle center-to-center distance and $\epsilon$ sets the unit of energy. 

The crossed orthogonal external fields induce orthogonal dipole 
moments $\bm{\mu}^m=\mu\mathbf{e}_m$ and $\bm{\mu}^e=\mu\mathbf{e}_e$ which we term for simplicity
as 'magnetic' and 'electric' dipoles (although the model is also appropriate for two electric moments). 
The coordinate frame is adjusted to coincide with the directions of these moments so that $\mathbf{e}_{m}=\mathbf{e}_{x}$ and $\mathbf{e}_{e}=\mathbf{e}_{y}$. 
In general these moments  could have different absolute values but for simplicity they are assumed to be 
equal. The two types of dipole moments are also assumed to be independent from each other and
interact only with dipole moments of the same type on other particles. 
Intuitively, one would model the interaction energy between dipoles of particles 1 and 2 by the point-dipole potential
\begin{equation}
U^{\alpha}_{dip}(\mathbf{r}_{12})=\frac{ \bm{\mu}^{\alpha}_1\cdot\bm{\mu}^{\alpha}_2}
{r_{12}^3}-3\frac{ (\bm{\mu}^{\alpha}_1\cdot\mathbf{r}_{12})(\bm{\mu}^{\alpha}_2\cdot\mathbf{r}_{12}) }{r_{12}^5}, 
\end{equation}
where $\alpha$ indicates the dipole type as being either $e$ or $m$. 
Due to the constraint $\bm{\mu}_1^{\alpha} \parallel \bm{\mu}_2^{\alpha}$ it follows that
\begin{equation}
\label{eq:2_0}
U^{\alpha}_{dip}(\mathbf{r}_{12})=
\frac{ \mu^{\alpha}_1\mu^{\alpha}_2}
{r_{12}^3}(1-3\frac{(\mathbf{r}_{12}\cdot\mathbf{e}_{\alpha})^2}{r_{12}^2}). 
\end{equation}
The resulting total dipolar interaction between two particles is the sum
of the dipolar potentials stemming from the magnetic and electric
dipoles, respectively.
Using $\mu=|\bm{\mu}^{\alpha}_i|$ and the relation $(\mathbf{r}_{12}\cdot\mathbf{e}_{e}+\mathbf{r}_{12}\cdot\mathbf{e}_{m})^2=r^2_{12}$ (which holds since
$\bm{\mu}^e$ and $\bm{\mu}^m$ are orthogonal) we obtain
\begin{equation}
\label{eq:2}
U^e_{dip}(\mathbf{r}_{12})+U^m_{dip}(\mathbf{r}_{12})=
-\frac{\mu^2}{r_{12}^3}. 
\end{equation}
The resulting interaction on the right side of Eq.~(\ref{eq:2}) is an isotropic, 
purely attractive interaction that lacks any kind of directional character. 
Therefore, the potential defined in Eq.~(\ref{eq:2}) can not generate any rectangular 
structures as observed in experiments~\cite{gangwal:nets, Biswal2013-magneticfractal}.
Moreover, having in mind that the field-induced spatial separation of charges in 
the ionic layer of a suspended particle is comparable to the particle diameter $\sigma$, 
a point dipole model seems even more unreasonable.

We therefore introduce an alternative model. 
Each dipole moment $\bm{\mu}^{\alpha}$ (with $\alpha=e,m$) is replaced by two opposite charges $-q^{\alpha_1}=q^{\alpha_2}$ 
which are shifted out of the particle center by a vector $\bm{\delta}^{\alpha_k}=(-1)^k\delta\bm{e_{\alpha}}$, with $k=1,2$. 
The vector $\bm{\delta}^{\alpha_k}$ points either parallel ($k=2$) or antiparallel ($k=1$) along the corresponding point 
dipole moment $\bm{\mu}^\alpha$. Independent of their shift or type, we set all charges 
to the same absolute value $q=|q^{\alpha_k}|=2.5(\epsilon/\sigma)^{-1/2}$.
The choice of the value $2.5$ is essentially arbitrary, as we will later normalize the interaction energy to eliminate the
dependence of its magnitude on the charge separation $\delta$ [see Eq.~(\ref{eq:u}) below]. 
Charges $k$ and $l$ on different particles i and j interact via a Yukawa potential
\begin{equation}
\label{eq:YU}
 U_{ij}^{\alpha_k\alpha_l}(r_{ij}) =  -q^2\frac{exp(-\kappa r^{\alpha_k\alpha_l}_{ij})}{r^{\alpha_k\alpha_l}_{ij}} 
\end{equation}
with  $r_{ij}^{\alpha_k\alpha_l}=|\mathbf{r}_{j}-\mathbf{r}_{i}+\bm{\delta}^{\alpha_l}-\bm{\delta}^{\alpha_k}|$. 
The inverse screening length is choosen to $\kappa=4.0\sigma^{-1}$ and a radial cutoff $r_c=4.0\sigma$ is applied. 
A schematic representation of the model with its internal arrangement of 'charges' is shown in Fig.~\ref{fig:1}. 
Due to the Yukawa-like interaction 
\begin{figure}[]
\centering
\includegraphics[width=0.5\textwidth]{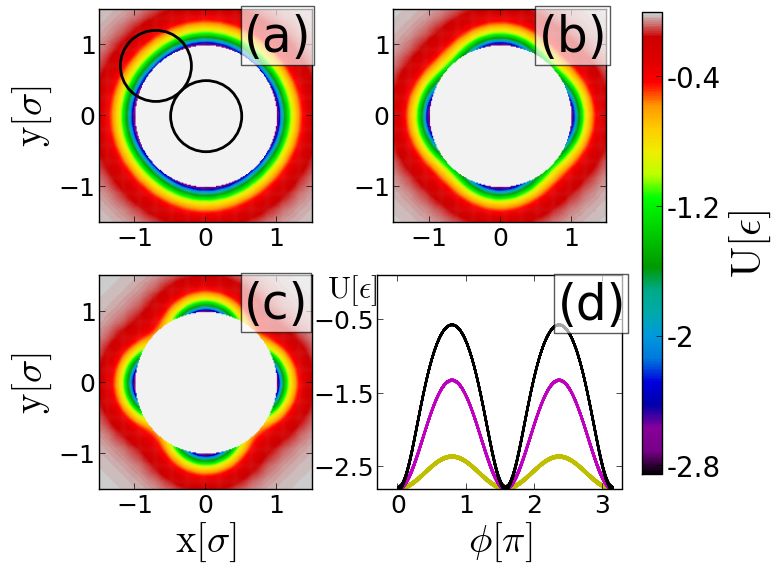}
\caption{\label{fig:2}(Color online) Normalized direction-dependent pair interaction $U(\mathbf{r}_{ij})$ [see Eq.~(\ref{eq:U})] between a particle in the center of the 
coordinate frame and a second particle (indicated as black circle in (a)) at various positions $\mathbf{r}_{ij}$ for three different 
charge separations $\delta=0.1,0.21,0.3\sigma$ corresponding to (a), (b), (c). Sterically excluded areas are indicated by white circles. 
(d) Interaction energy at distance $r_{ij}=\sigma$ as function of $\phi$, 
the angle measured in multiples of $\pi$ against the x-axis, for $\delta=0.1\sigma$ (yellow), $\delta=0.21\sigma$ (purple) and $\delta=0.3\sigma$ (black).}
\end{figure}
our model lacks the long-range character of true dipolar interactions. 
However, similar models with comparable interaction ranges have been used to describe 
dipolar colloids in the context of discontinous molecular dynamics simulations~\cite{Schmidle2012-dipolegel,Goyal2008-dipole}. 
We note that mimicking magnetic dipoles via spatially separated 'charges' appears somewhat artificial 
from a physical point of view. 
However, here we could think of a particle with a strongly inhomogenous distribution of magnetic moments.
Moreover, the idea behind our {\em ansatz} is not to mimick a particular complex colloid, 
but rather to provide a generic model for field-induced directional interactions.

The arrangment of charges inside particles then results in a pair-interaction $U_{DIP}(\mathbf{r}_{ij})$ given by
\begin{equation}
\label{eq:Unon}
U_{DIP}(\mathbf{r}_{ij})=\sum^2_{k,l=1}[U^{e_ke_l}_{ij}(\mathbf{r}_{ij})+U^{m_km_l}_{ij}(\mathbf{r}_{ij})].
\end{equation}
In principle, $U_{DIP}(\mathbf{r}_{ij})$ is a function of $q$ and $\delta$. 
To facilitate the comparison between the interactions at different $\delta$ ($q$ is chosen to be constant), 
we normalize $U_{DIP}(\mathbf{r}_{ij})$ according to %by a function $f(\delta)$ 
\begin{equation}
\label{eq:normalize}
\tilde{U}_{DIP}(\mathbf{r}_{ij}) = U_{DIP}(\mathbf{r}_{ij}) \times u/U_{DIP}(\sigma \mathbf{e}_{\alpha})
\end{equation}
where the constant $u=-2.804\epsilon$ is calculated from the unnormalized energy $U_{DIP}(\sigma\mathbf{e}_{\alpha})$ 
with model parameters $\delta=0.3\sigma$ and $q=2.5(\epsilon/\sigma)^{-1/2}$.
This procedure ensures that the normalized energy between two particles
at contact ($r_{ij}=\sigma$) and direction $\mathbf{r}_{ij}=\sigma\mathbf{e}_{\alpha}$ (pointing along one of the fields) 
has the constant value $u$ for all $\delta$, that is 
\begin{equation}
\label{eq:u}
\tilde{U}_{DIP}(\sigma\mathbf{e}_{\alpha})=u. 
\end{equation} 
The full pair interaction of our model is then given by
\begin{equation}
\label{eq:U}
U(\mathbf{r}_{ij})=U_{SS}(\mathbf{r}_{ij})+\tilde{U}_{DIP}(\mathbf{r}_{ij}).
\end{equation}
The resulting potential is illustrated in Fig.~\ref{fig:2}(a)-(c) for a particle in the center of the coordinate frame and a
second particle at various distances $r_{ij}$ and angles $\phi=\arccos(\mathbf{r}_{12}\cdot\mathbf{e}_x/r_{12})$
with 'charge' separations $\delta=0.1,0.21,0.3\sigma$. 
The value $\delta=0.21\sigma$ is motivated by our simulation results presented in Sec.~\ref{Local Order}.
Sterically-excluded areas are shown in white and energy values are color coded in units of $\epsilon$.
The weak anisotropy of the resulting particle interactions at small $\delta$ (where one essentially adds two dipolar potentials,
see Eq.~(\ref{eq:2})) transforms to a patchy-like pattern by increasing $\delta$.
Energy minima become more and more locally restricted and particle interactions become directional in character.
This is also seen in Fig.~\ref{fig:2}(d) which gives the energy between two particles in contact as function of $\phi$ for different $\delta$.
From Fig.~\ref{fig:2}(d) we also see that, independent of the 'charge' separation $\delta$, the minima of the full interaction potential [see Eq.~(\ref{eq:U})] 
occur for connection vectors $\mathbf{r}_{ij}=\sigma\mathbf{e}_{e}$ and $\mathbf{r}_{ij}=\sigma\mathbf{e}_{m}$ (i.e., pointing along the fields).
Note that this already holds for the unnormalized energy given in. Eq.~(\ref{eq:Unon}).
\begin{figure*}
\centering
\includegraphics[width=0.99\textwidth]{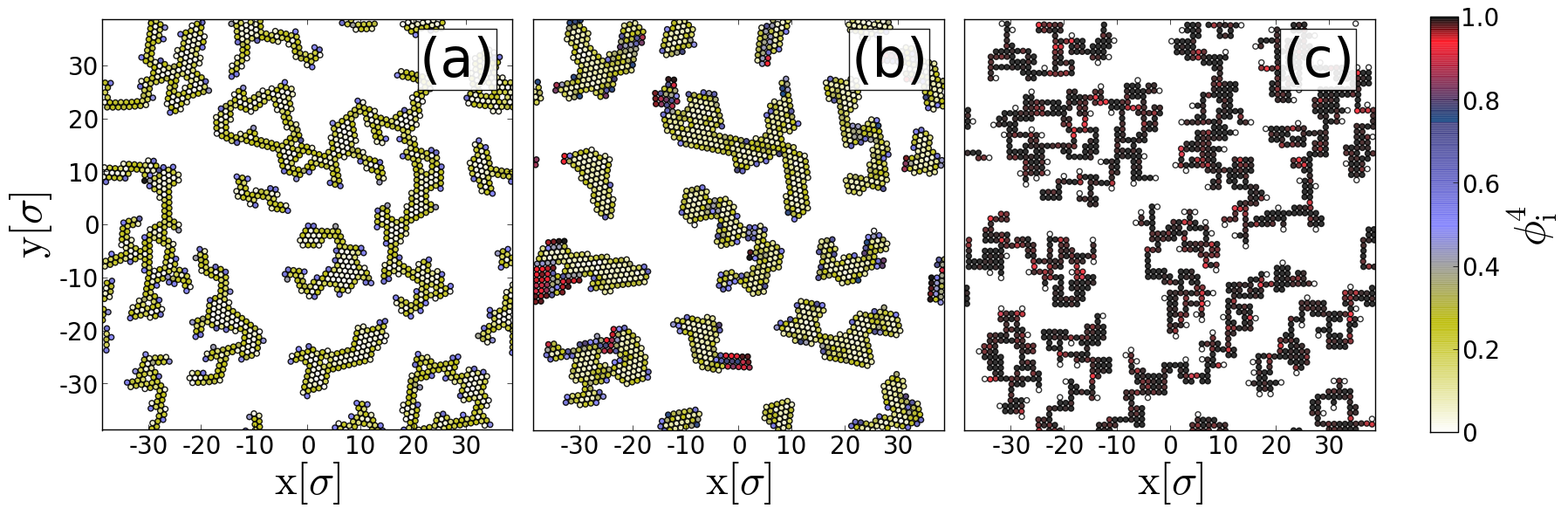}
\caption{\label{fig:3}(Color online)  Simulation snapshots at $\rho^*=0.3$ and $T^*=0.05$ for 
(a) $\delta=0.1\sigma$, (b) $\delta=0.21\sigma$ and (c) $\delta=0.3\sigma$. Particles are colored according to their value of $\phi_i^4$. }
\end{figure*}
Simulations are performed with $N=1800$ to $3200$ particles at a range 
of reduced number densities $\rho^*=\rho\sigma^2$ and temperatures $T^*=k_BT/\epsilon$, 
in a square-shaped simulation cell with periodic boundary conditions. 
The equations of motion
\begin{equation}
\dot {\mathbf r}_{i} = \frac{T^*}{\gamma}\sum^N_{j=1} \bm \nabla U(\mathbf{r}_{ij})+ \bm \zeta_i(T^*)
\label{eq:dyn}
\end{equation}
are solved via the Euler scheme with an integration stepwidth $\Delta t = 10^{-4}\tau_b$, 
where $\tau_b$ is the Brownian timescale defined by $\tau_b=\sigma^2\gamma/T^*$, $\gamma=1.0$ is a friction constant 
and $\bm \zeta_i(T^*)$ is a Gaussian noise vector which acts on particle $i$ and fulfills the relations
$\langle \bm \zeta_i \rangle = 0$ and 
$\langle \bm \zeta_i (t) \bm \zeta_j (t') \rangle = 2 \gamma k_BT \delta_{ij}\delta(t-t')$~\cite{Ermak1975}. 
We perform simulations for up to $10^3\tau_b$.

\subsection{Target quantities}
To characterize the structure of the systems we consider several quantities. The first one is the mean coordination number
\begin{equation}
\bar{z}=\frac{1}{N}\sum_{i=1}^N{z_i},
\end{equation}
where $z_i$ is the number of neighbors of particle $i$ and the sum is over all particles. 
In the following, two particles are considered to be nearest neighbors if their center-to-center distance is smaller than $r_b=1.15\sigma$. 

To identify local particle arrangements, the orientational bond order parameter is of special importance. 
For particle $k$ it is given by
\begin{equation}
\label{eq:phi}
\phi_k^n=\frac{1}{z_k}\sum^{z_k}_{l=1} |\exp(in\theta^{kl}_{\lambda})|
\end{equation}
with $z_k$ being the number of neighbors and $\theta^{kl}_{\lambda} = \arccos(\mathbf{r}_{kl}\cdot\mathbf{r}_{k\lambda}/({r}_{kl}{r}_{k\lambda}))$
being the angle between the bond of particle $k$ and its neighbor $l$ measured 
against a randomly chosen bond of particle $k$ to one of its neighbouring particles $\lambda$. Hence, $\phi_k^n=0$ for $z_k<2$.
The integer value $n$ determines the type of order 
which is detected by this parameter. We concentrate on $\phi_4$ and $\phi_6$ to identify square (rectangular) and hexagonal lattice types.
Its ensemble average is calculated via
\begin{equation}
\Phi_n=\frac{1}{N}\sum_{i=1}^N\phi^n_i.
\end{equation} 
The reversibility of 'bond' formation and slipperyness of existing bonds can be characterized by the bond and 
the bond-angle auto-correlation functions $c_b(t)$ and $c_a(t)$. 
To evaluate $c_b(t)$ we assign a variable $b_{ij}(t)$ to each pair of particles at each time step 
which is $1$ if the particles $i$ and $j$ are nearest neighbors or zero otherwise. The bond auto-correlation function is then defined as 
\begin{equation}
c_b(t)=\langle b_{ij}(t_0)b_{ij}(t)\rangle, 
\end{equation}
where the brackets indicate an average over all pairs that are bonded at time $t_0$. 
The bond-angle auto-correlation function $c_a(t)$ is defined similarly by defining the unit vector
\begin{equation}
\mathbf{a}_{ij}(t) = \mathbf{r}_{ij}(t)/r_{ij}(t),
\end{equation}
such that
\begin{equation}
c_a(t)=\langle 1-\arccos{(\mathbf{a}_{ij}(t)\cdot\mathbf{a}_{ij}(t_0))} / \pi\rangle
\end{equation}
where we average again over all pairs. While $c_b$ gives the information on how stable bonds are over time, $c_a$ 
tells how stable their direction is over time. 
Note that in contrast to the typical definition of correlation functions for stationary systems~\cite{hanson},
here the functions $c_b(t)$ and $c_a(t)$ are not independent of the time origin $t_0$.

Finally, we consider the fractal dimension $D_f$ of particle clusters, which is particularly important in the context of DLCA. 
Clusters are defined as a set of particles with common next neighbors. 
The size of a cluster is then quantified by its radius of gyration 
\begin{equation}
R^2_g=\frac{1}{N_{cl}}\sum^{N_{cl}}_{i=1}(\mathbf{r}_{i}-\mathbf{\bar{r}})^2,
\end{equation}
where $N_{cl}$  is the number of particles in the cluster, and  $\mathbf{\bar{r}}$ is the postion of its center-of-mass.
By plotting $\ln{R_g}$ against $\ln N_{cl}$ for  different clusters, 
we extract the fractal dimension $D_f$ via the relationship $R_g \sim N^{1/D_f}_{cl}$ (see Ref.~\cite{Witten1981-DLA,Biswal2013-magneticfractal}). 

\section{Results}
\label{Results}
Our large-scale Brownian dynamics simulations show that the system is very sensitive to changes in temperature $T^*$, 
number density $\rho^*$, and charge separation $\delta$. 
In this large parameter space we find a variety of different states ranging from small fractal aggregates and single-chain structures
at low temperatures to coarser, 
isolated or interconnected clusters at higher temperatures. In the following sections \ref{Local Order}~-~\ref{Diffusion limited aggregation} 
we first discuss the structure, the time correlation functions and the fractal dimensions at a low temperature and an intermediate density, 
focussing on the impact of the model parameter $\delta$. In section~~\ref{Beyond DLCA - Higher Temperatures} and~~\ref{Spotlight on higher densities} 
we then turn to the impact of temperature and density.

\subsection{Effect of Charge Separation on Local Order }
\label{Local Order}

\begin{figure}[]
\centering
\includegraphics[width=0.34\textwidth]{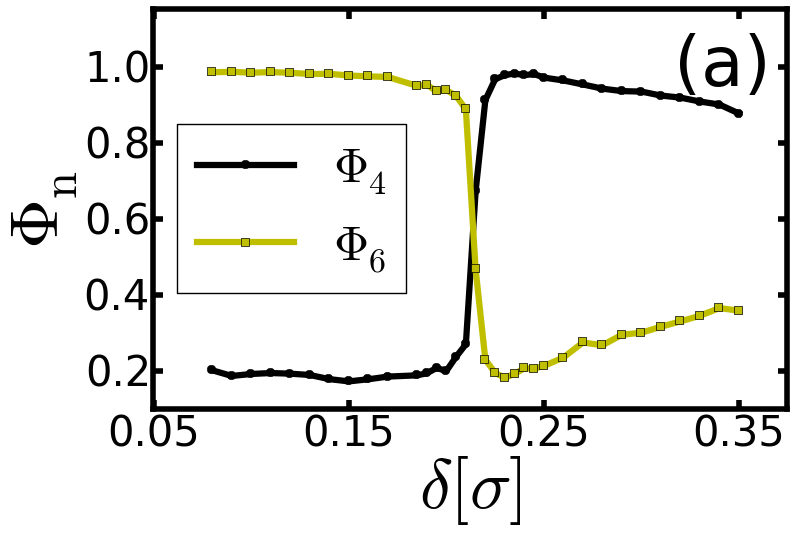}
\includegraphics[width=0.34\textwidth]{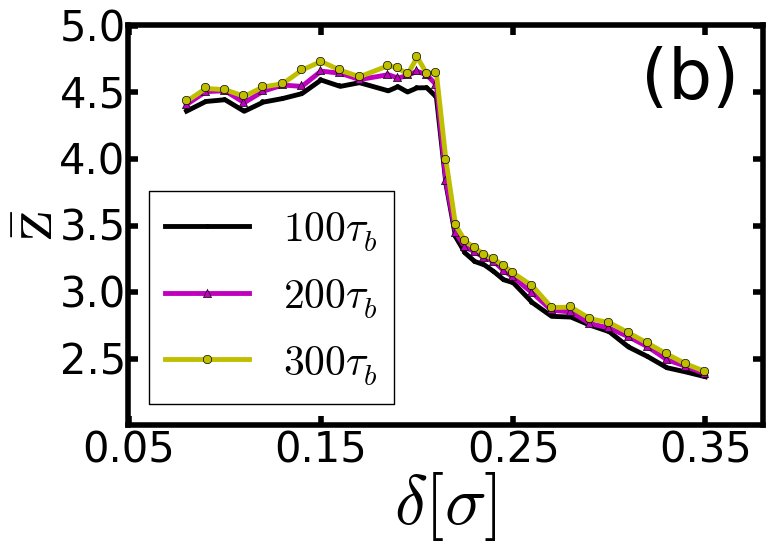}
\caption{\label{fig:4}(Color online)  Results for simulations with $N=1800$ at temperature $T^*=0.05$ and density $\rho^*=0.3$. 
(a) Orientational bond order parameters $\Phi_4$ for square (black) and $\Phi_6$ (yellow) for hexagonal particle arrangements. 
(b) Mean coordination number $\bar{z}$ as function of charge separation $\delta$ at times $t=100,200,300 \tau_b$. }
\end{figure}

At first we study the system at low temperature $T^*=0.05$ and intermediate density $\rho^*=0.3$ for different charge separations $\delta$. 
In Fig.~\ref{fig:3} simulation snapshots for $\delta=0.1\sigma,0.21\sigma,0.3\sigma$ at $t=300\tau_b$ [see. Eq.~(\ref{eq:dyn}) below] are 
shown, where $\tau_b$ is the Brownian timescale.
The colorcode reflects the orientational bond order parameter $\phi^4_i$ of each particle $i$. 
All three cases are characterized by clusters with irregular shapes. However, local particle arrangements differ strongly. 
While for $\delta=0.1\sigma$ the particles aggregate in a hexagonal fashion, at $\delta=0.3\sigma$ they aggregate into rectangular structures. 
At the intermediate charge separation $\delta=0.21\sigma$, hexagonal order dominates the system; 
however, some clusters also reveal subsets of particles in rectangular arrangements.
A more quantitative description is given by the orientational bond order parameters $\Phi_{4(6)}$ shown 
in Fig.~\ref{fig:4}(a) as functions of $\delta$. By increasing $\delta$, one observes a sharp transition at $\delta \approx 0.21\sigma$ 
from hexagonal towards rectangular (square) order. 

\begin{figure}[]
\centering
\includegraphics[width=0.35\textwidth]{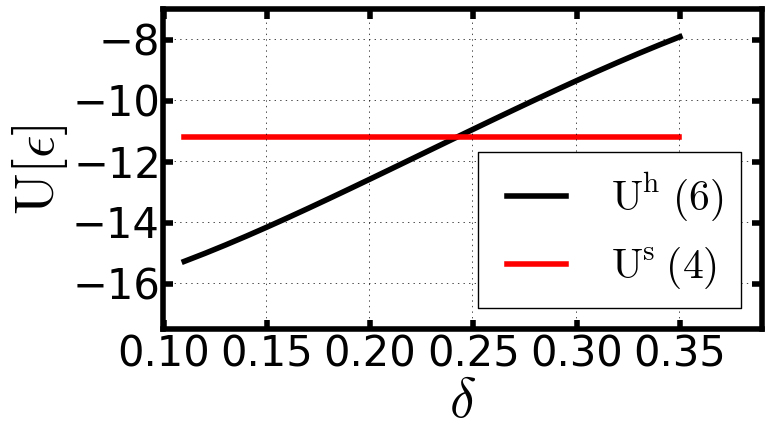}
\caption{\label{fig:6_2}(Color online)  Minimum energy of a particle with six neighbors in hexagonal arrangement as function of $\delta$ (black) 
and energy for a particle in rectangular arrangement with 4 neighbors (red).}
\end{figure}

The very presence of such a sharp transition can be explained via energy arguments 
based on the $\delta$-dependent pair potential plotted in Figs.~\ref{fig:2}(a)-(d).
To this end, we calculate the energy $U_i^{hex}(\delta)=\sum_{j=1}^6 U({\bf r}_{ij})$
of a particle $i$ with six neighbors $j$, which are located in a hexagonal arrangement
at 'contact' distance $\sigma$ around $i$.
Note that not all hexagonal configurations do have the same contact energy. 
This is due to the anisotropy of interactions, see Fig.~\ref{fig:2}(d). 
Therefore we consider a hexagonal configuration in which 
the contact energy is as low as possible (this configuration was found numerically).
The dependence of this lowest contact energy $U_i^{hex}(\delta)$ on the charge separation
parameter is plotted in Fig.~\ref{fig:6_2}. Also shown is the corresponding 
energy $U_i^{sq}(\delta)=\sum_{j=1}^4 U({\bf r}_{ij})=4\times u$
of a particle with four neighbors $j$ located at distance $\sigma$ in a rectangular 
arrangement, i.e., in the energy minima around $i$ (the quantity $u$ was defined
below Eq.~(\ref{eq:normalize})). 
Note that the energy $U_i^{sq}(\delta)$ does
not depend on $\delta$ according to Eq.~(\ref{eq:u}). 
As shown in Fig.~\ref{fig:6_2}, 
the two curves intersect at a "critical" value of $\delta=0.24\sigma$. 
Thus, the simple energy arguments already suggest a transition 
between states with local
hexagonal and square order, even though the predicted critical value 
is somewhat larger than the value of $\delta=0.21\sigma$ seen in the actual simulations
at finite temperature and density [see Fig.~\ref{fig:4}(a)].

Further information is gained from the behavior of the mean coordination number as a function of $\delta$ plotted in  
Fig.~\ref{fig:4}(b) for three different times $t=100\tau_b$, $200\tau_b$ and $300\tau_b$.
At all times considered, $\bar{z}$ undergoes a steep decrease at $\delta \approx 0.21\sigma$ 
from a nearly constant value, $\bar{z}_{hex} \approx 4.5$, to a value $\bar{z}_{sq}\approx 3.5$. 
This behavior reflects, on the one hand, again the presence of a sharp transition; on the other hand, the actual values of $\bar{z}_{hex}(\bar{z}_{sq})$ reveal
the "non-ideal" character of the aggregates in terms of coordination numbers. 
For example, for $\delta > 0.21\sigma$ we find that $\bar{z}$ and $\Phi_{4}$ decrease with $\delta$, while $\Phi_{6}$ increases.
However, this does not indicate a decline of the rectangular order; it rather results from an increasing 
amount of particles residing in chains oriented either in x- or y-direction. 
The coordination number $z_i$ of a particle $i$ in such a chain is $\leq 2$, leading to a mean coordination number $\bar{z}<4$. 
Furthermore, the parameters $\phi_i^4$ and $\phi_i^6$ 
[see Eq.~(\ref{eq:phi})] become unity for a particle forming exactly two bonds under an angle of $\pi$ (straight chain). 
This does not affect $\Phi_4$, which is already large at $\delta>0.21\sigma$, but significantly increases $\Phi_6$.
Finally, the counter-intuitive decrease of $\Phi_4$ with $\delta$ results from the increasing amount of particles 
with only one neighbor (e.g., ends of chains appearing white in  Fig.~\ref{fig:3}(c)). 
These particles yield no contribution to $\Phi_4$ [see Eq.~(\ref{eq:phi})].

The "non-ideal" values of $\bar{z}_{hex}$ and $\bar{z}_{sq}$ also explain why our energy argument for the location
of the hexagonal-to-square transition, which was based on ideal arrangements with six and four neighbors, respectively, 
does yield the transition value 
$\delta=0.24\sigma$ rather than $\delta=0.21\sigma$ obtained from simulation.
We can now reformulate the argument by using
the actual mean coordination numbers extracted from 
our simulations, $\bar{z}_{sq}=3.5$ (instead of $4$) and $\bar{z}_{hex}=4.5$ (instead of $6$).
Following the calculations for the ideal arrangements described before, the energy of the square-like arrangement
is $U^{sq}=3.5\times u$. For the hexagonal arrangement, we use the average minimum energy with either $z_i=4$ or $z_i=5$ neighbors,
yielding $\bar{U}^h(\bar{z}_{hex},\delta)=(U^{hex}(4,\delta)+U^{hex}(5,\delta))/2$.
The resulting critical value of the charge separation is
$\delta \approx 0.21\sigma$, which coincides nicely with the transition value observed in our simulations.

\subsection{Transient character of aggregates}
\label{Transient character of aggregates}

Although the local structures characterized by $\bar{z}$ and $\Phi_{4(6)}$ persist, in general, over the simulation times considered,
we are still facing a {\em transient} (out-of-equilibrium) structure formation as seen, e.g., from the slight increase of $\bar{z}$ with time in Fig.~\ref{fig:4}(b).
This raises a question about the typical "lifetime" of the aggregates.  

To this end we now consider dynamical properties, 
namely the bond and bond-angle auto-correlation functions, $c_b(t)$ and $c_a(t)$. 
It is not reasonable to extract decay rates from these functions (as it is usually done) because 
in transient states, decay rates are, strictly speaking, functions of time themselves. Still, it is interesting to see whether the temporal 
correlation of bonds (bond angles) for different $\delta$ allows us to distinguish between {\em qualitatively} different aggregation regimes. 
\begin{figure}[]
\centering
\includegraphics[width=0.49\textwidth]{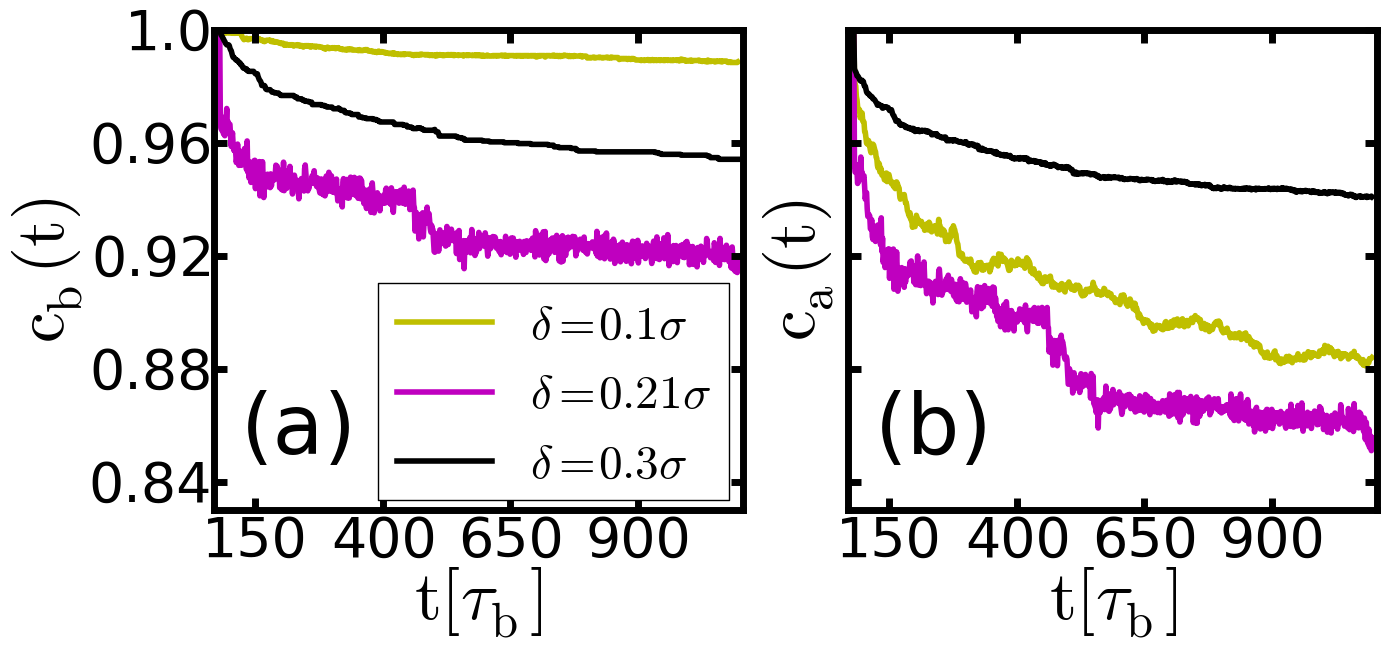}
\caption{\label{fig:5}(Color online)  Time correlation functions obtained from simulations with $N=1800$ at temperature $T^*=0.05$ and density $\rho^*=0.3$. 
(a) [(b)] Time evolution of the bond [angle] autocorrelation function  $c_b(t)$ [$c_a(t)$] for three different charge separations 
$\delta = 0.1\sigma, 0.21\sigma$ and $0.3 \sigma$ colored in yellow, purple and black respectively.}
\end{figure}

Numerical results for $c_b(t)$ and $c_a(t)$ are plotted in Figs.~\ref{fig:5}(a) and (b), respectively, where we consider a large time range
up to $t\approx 10^3\tau_b$. The time axis starts at the finite time when all the systems have formed stable aggregates.
The data in Figs.~\ref{fig:5}(a) and (b) pertain to three representative values of the charge separation parameter related to the hexagonal structures ($\delta=0.1\sigma$), 
rectangular structures ($\delta=0.3\sigma$), and to the transition region ($\delta=0.21\sigma$).
In the square regime ($\delta=0.3\sigma$) the decay of both $c_b(t)$ and $c_a(t)$ is almost identical and very slow.
From this we conclude that the square regime is characterized by almost unbreakable bonds with {\em fixed} orientations. 
This is different in the hexagonal regime ($\delta = 0.1\sigma$) where $c_b(t)$ remains 
nearly constant even after long times (meaning that bond-breaking is very unlikely), while $c_a(t)$ decays
much faster. Thus, the direction of bonds are less restricted. We interprete this behavior as evidence
that two particles, though being bonded, are still able to rotate around each other to some extent. This is a characteristic feature 
of {\em slippery} bonds. 
Finally, in the transition regime ($\delta=0.21\sigma$) both functions $c_b(t)$ and $c_a(t)$ decay significantly faster than in the other cases, 
with the decay of the bond-angle correlation function being even more pronounced.
In that sense we may consider the bonds in the transition region also as slippery (although less long-lived than in the other cases).

We conclude that the different structural regimes identified in the preceding section are indeed characterized by different relaxational dynamics.
Moreover, all of the observed aggregates have lifetimes of at least several hundered $\tau_b$.
Such long-lived bonds are indicative of diffusion limited cluster-cluster aggregation. In the next section
we therefore consider the fractal dimension.

\subsection{Diffusion limited aggregation}
\label{Diffusion limited aggregation}
\begin{figure}[]
\centering
\includegraphics[width=0.35\textwidth]{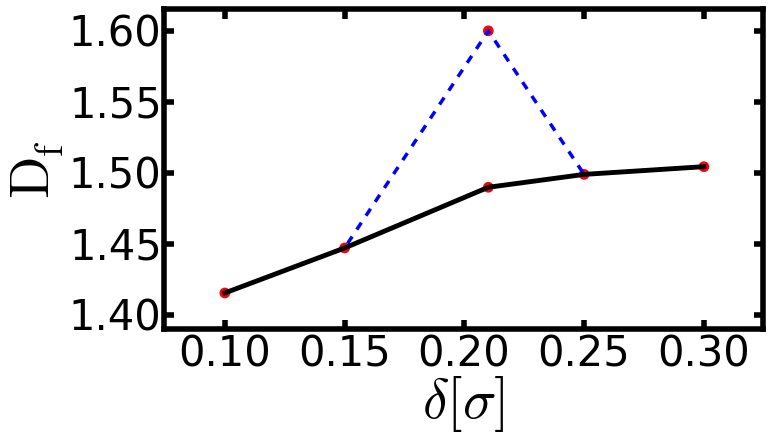}
\caption{\label{fig:6}(Color online)  Fractal dimension $D_f$ as a function of charge separation at $\rho^*=0.3$ and $T^*=0.05$. 
At $\delta=0.21\sigma$  we find a bimodal distribution of fractal  dimension with peaks at $D_f=1.48$ (solid line) and $1.6$ (dashed line).}
\end{figure}
In Fig.~\ref{fig:6} the fractal dimension $D_f$ is shown as a function of $\delta$ at time $t=250\tau_b$, 
density $\rho^*=0.3$ and temperature $T^*=0.05$. We find that $D_f$ increases slightly 
with $\delta$ but remains in a range between $1.4$ and $1.5$, except at $\delta=0.21\sigma$. 
There, the fractal dimension exhibits a bimodal distribution, taking values between $D_f \approx 1.48$ and $D_f\approx 1.6$
(dashed line in Fig.~\ref{fig:6}).

Despite these variations, the values of $D_f$ found here are significantly 
smaller than the fractal dimension $D_f=1.71$ observed in earlier studies of DLCA in 
two-dimensional continuous (off-lattice) systems~\cite{Witten1981-DLA,Meakin1989-DLA2D}. 
Except for the case $\delta=0.21\sigma$, the values 
in Fig.~\ref{fig:6} are comparable with previous findings for DLCA in two-dimensional {\em lattice} systems 
and systems with spatial or interaction anisotropies~\cite{Popescu2004-anisotropicDLA,Ball1986-anisotropicDLA,Lee1996-tempDLA2}.  
The present system is indeed anisotropic in the sense that 
the external fields impose preferences on the directions of particle bonds and therefore also on the orientations of aggregates. 
This effect is most pronounced in the rectangular regime ($\delta=0.3\sigma$).
Therefore, it is plausible that our system undergoes a special case of anisotropic DLCA, in (quantitative) 
accordance with experimental results~\cite{Biswal2013-magneticfractal} and 
theoretical predictions~\cite{Meakin1995-DLAreview,Popescu2004-anisotropicDLA,Ball1986-anisotropicDLA,Helgesen1988-dipoleDLA}. 
We should note that, due to our simulation method, the cluster sizes (typically 
involving $10^1-10^3$ particles) are relatively small compared to the particle numbers 
considered in the literature ($10^6$ particles)~\cite{Meakin1989-DLA2D,Kuijpers2013-DLAoptimization,Meakin1995-DLAreview}. 
A more detailed analysis of the impact of anisotropic interactions on the fractal structure is beyond the scope of this study. 

We also relate our findings to the newer concept of slippery DLCA~\cite{Nicolai2008-slippery,Mason2007-slippery}, 
where the bonds are essentially unbreakable but able to rotate. 
Indeed, as discussed in section~\ref{Transient character of aggregates}, bonds are slippery in nature for
small $\delta$ in the hexagonal regime. For three-dimensional systems it has been reported \cite{Nicolai2008-slippery,Mason2007-slippery}
that the fractal dimension $D_f$ remains the same for slippery and classical DLCA, while the mean coordination number 
$\bar{z}$ differs. Specifically,  $\bar{z}$ is significantly higher for sDLCA~\cite{Nicolai2008-slippery,Mason2007-slippery}. 
The same observation emerges when we consider our values of $\bar{z}$ plotted in Fig.~\ref{fig:4}(b), from which 
one sees a pronounced decrease of $\bar{z}$ upon entering the square (DLCA) regime.
However, in contrast to earlier studies we find
$D_f$ to slightly increase with $\delta$, especially in the hexagonal regime. 
We interpret this behavior as a consequence of the fact 
that binding energies in the hexagonal regime decrease with increasing values of $\delta$, 
while they remain constant in the square regime (see Fig.~\ref{fig:6_2}).
The corresponding stability of bonds should be correlated to the binding energies
which explains the slightly increasing values of $D_f$ in the hexagonal regime.
\begin{figure*}[]
\centering
\includegraphics[width=0.3\textwidth]{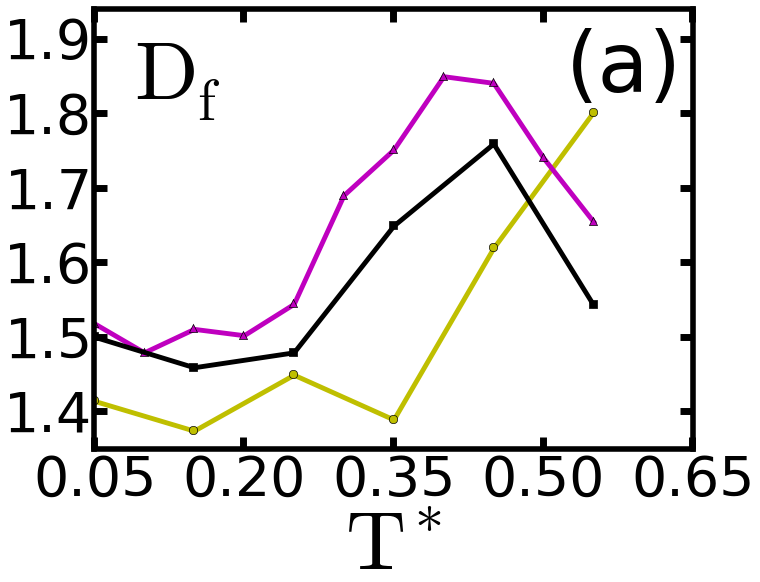}\includegraphics[width=0.3\textwidth]{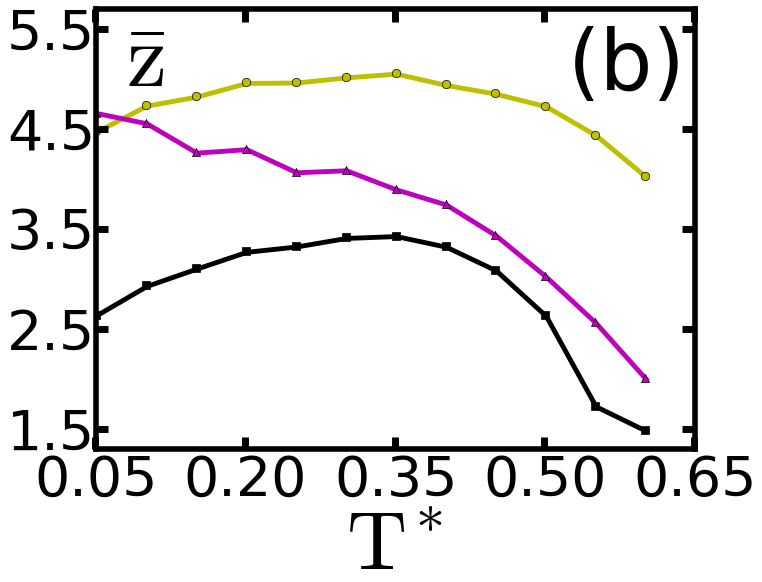}\includegraphics[width=0.3\textwidth]{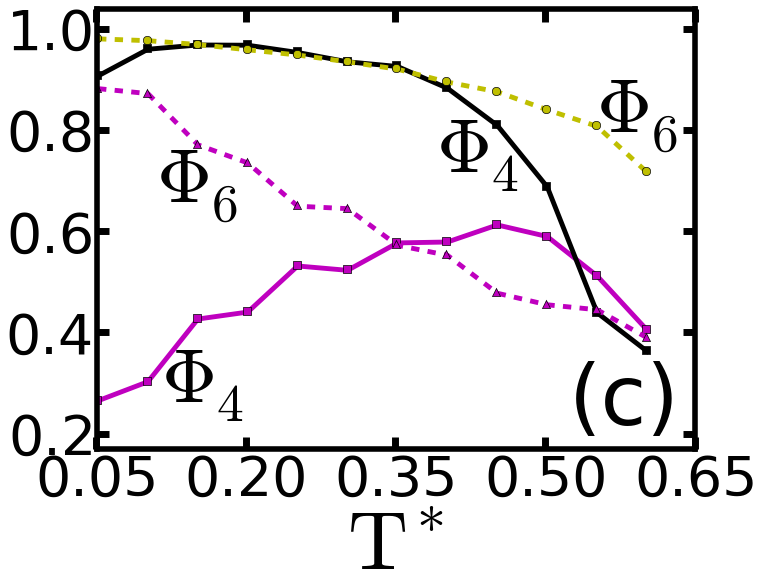}
\caption{\label{fig:7}(Color online)  Temperature dependence of the system properties at density $\rho^*=0.3$ for charge separations $\delta=0.1,0.21,0.3\sigma$ 
colored in yellow, purple and black, respectively. (a) Fractal dimension $D_f$ evaluated at $t\approx250\tau_b$, (b) Mean coordination number, 
(c) Orientational order parameters $\Phi_4$ and $\Phi_6$.
}
\end{figure*}

Finally, in the transition region ($\delta = 0.21\sigma$) we found a bimodal
distribution of the fractal dimensions $D_f$ with maxima at $D_f\approx 1.48$ and
$D_f\approx 1.6$. This second maximum corresponds to only $\approx 25\%$ of the considered cases (twelve
independent simulation runs). The first maximum at $D_f\approx 1.48$ therefore
clearly dominates and fits nicely to the functional dependence of $D_f$ on $\delta$ (see Fig.~\ref{fig:6}). 
We assume that the less frequent peak results from a ’switching’ of the local structures between hexagonal
and rectangular arrangements, which is accompanied by a significantly larger bond-breaking probability (see Fig.~\ref{fig:5}(c)).
Again this allows compactification of aggregates and increases the fractal dimension in the transition regime.

\subsection{Beyond DLCA - Higher Temperatures}
\label{Beyond DLCA - Higher Temperatures}

Diffusion limited aggregation is restricted to systems with attractive particle interactions much stronger than $k_BT$. 
By increasing the temperature sufficiently, thermal fluctuations become able to break bonds which results in a faster decay 
of the the bond auto-correlation functions and a compactification of aggregates. 
Indeed, for square lattice models it was found that $D_f$ is a 
monotonically increasing function of temperature~\cite{Lee1996-tempDLA2,Ito2009-dipoleDLAtemp}.

In Fig.~\ref{fig:7}(a) the fractal dimension $D_f$ of the present model is plotted as a function of 
temperature $T^*$ for charge separations $\delta=0.1\sigma, 0.21\sigma$ and $0.3\sigma$. 

\begin{figure}[!b]
\centering
\includegraphics[width=0.33\textwidth]{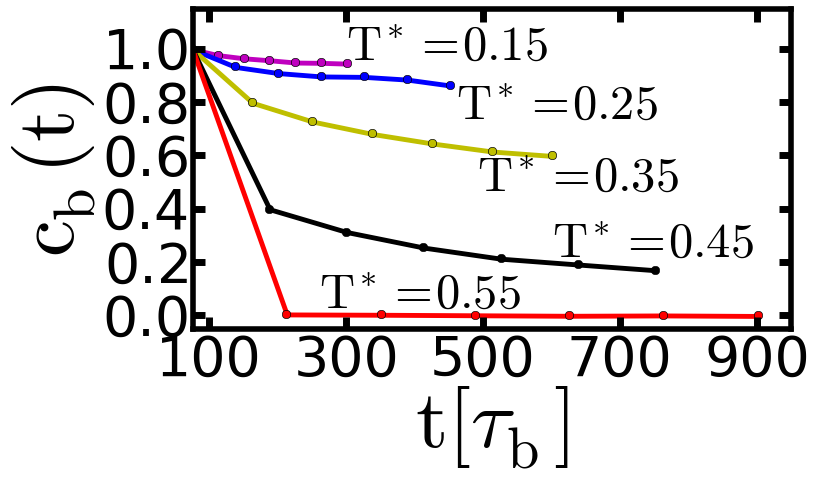}
\caption{\label{fig:cb_temp}(Color online)  Bond auto correlation function $c_b(t)$ for different 
temperatures $T^*$ at charge separation $\delta=0.3\sigma$ and density $\rho^*=0.3$.}
\end{figure}

We first concentrate on the case $\delta=0.3\sigma$, corresponding to the square regime at low $T^{*}$. 
In the range of very low temperatures $T^{*} < 0.25$, the fractal dimension
is small and stays essentially constant. Increasing $T^{*}$ towards slightly larger values then leads to an increase of $D_f$, reflecting the (expected) compactification.
This increase of $D_f$ is accompanied by an increase of the
mean coordination number $\bar{z}$ [see Fig.~\ref{fig:7}(b)] within the temperature range considered,
indicating the growing number of bonds due to local and global structural reconfigurations. 
The corresponding changes in the stability of the bonds are illustrated in Fig.~\ref{fig:cb_temp}, where we have plotted
the time evolution of $c_b(t)$ for several temperatures 
(at $\delta=0.3\sigma$). Clearly, the decay of $c_b(t)$ becomes faster for higher 
temperatures. This is the reason why structural reconfigurations and, in consequence, compactification of aggregates becomes possible.
\begin{figure*}[]
\centering
\includegraphics[width=0.99\textwidth]{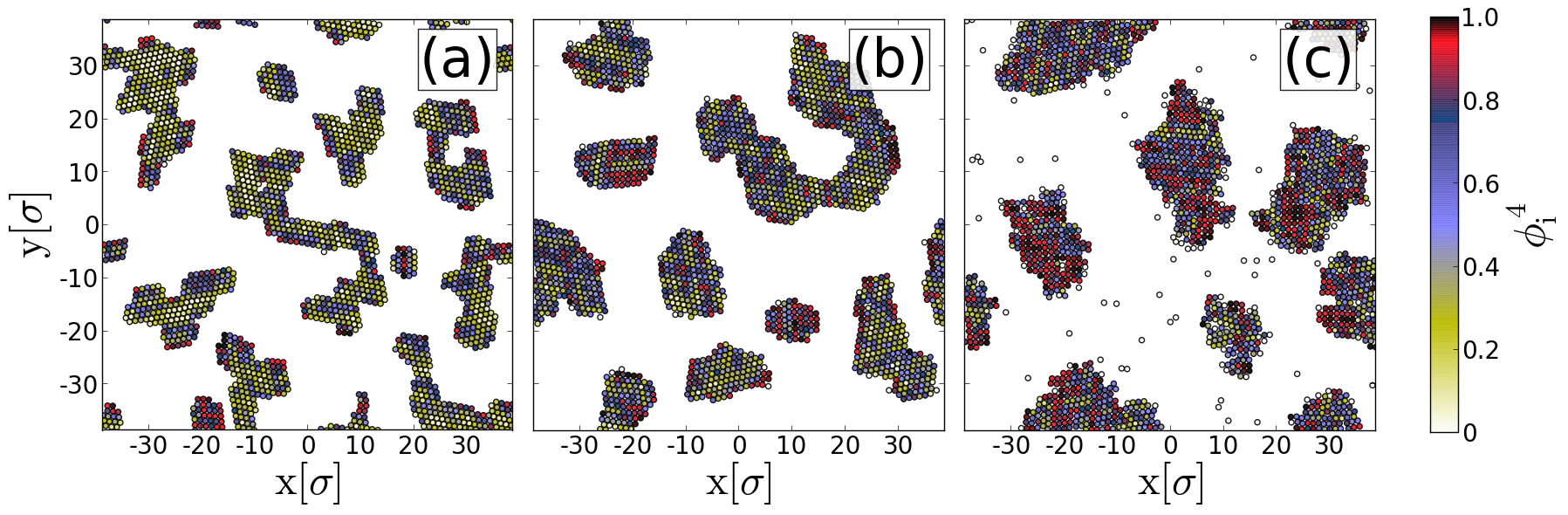}
\caption{\label{fig:8}(Color online)  Simulation snapshots at $\rho^*=0.3$ with $\delta=0.21\sigma$ at (a) $T^*=0.15$, (b) $T^*=0.3$, and (c) $T^*=0.45$. 
Particles are colored according to their value $\phi_i^4$. }
\end{figure*}
These trends persist until $T_{f,sq}^*\approx0.375$,
beyond which the system at $\delta=0.3\sigma$ starts to behave in a qualitatively different way. 
The mean coodination number $\bar{z}$ displays a maximum and subsequently a rapid decay. 
We also find that the fractal dimension has not yet reached its maximum value at  $T_{f,sq}^*$; this maximum occurs at the slightly larger temperature
$T^*\approx 0.42$ (see Fig.~\ref{fig:7}(a)). This 'delay' of $D_f$ can be understood from the fact that, upon the entrance of bond-breaking,
filigree parts of the 
aggregates are more likely affected than more compact ones. Hence, the fraction of 'compact' small aggregates still grows.
Even more important, the function $\Phi_4(T^*)$ in Fig.~\ref{fig:7}(c) displays a pronounced decay of rectangular order for $T^*>T_{f,sq}$.
From the sum of these indications we conclude that, at temperatures higher than  $T^*_{f,sq}\approx0.375$, 
the system transforms into a (stable or metastable) {\em fluid} phase.
In this fluid phase, the {\em overall} structure starts to become homogeneous and isotropic, while the local structures involve 
only a small number of bonds with short bond-life times.

For the system at $\delta=0.1\sigma$ (hexagonal structure at low $T^{*}$), an estimate of the
 "fluidization" temperature $T^*_{f,hex}$ based on the behavior of order parameters, coordination number and fractal dimension is more speculative.
Nevertheless, the data suggest that  $T^*_{f,hex}>T^*_{f,sq}$. This is indicated, first, by the fact that $\Phi_6(T^*)$ decays only very slowly with temperature
until $T^{*}\approx 0.6$ (see Fig.~\ref{fig:7}(c)). Second,  the mean coordination number shows only a weak maximum (and no fast decay afterwards) 
compared to the case $\delta=0.3\sigma$. Third, the fractal dimension keeps increasing with $T^{*}$ for all considered temperatures $T^*<0.6$.
Therefore we conclude that $T^*_{f,hex} > 0.6$.
We understand this higher fluidization temperature at $\delta=0.1\sigma$ from the fact that binding energies in hexagonal structures are larger; 
therefore, higher coupling energies must be overcome.

\begin{figure}[!b]
\centering
\includegraphics[width=0.3\textwidth]{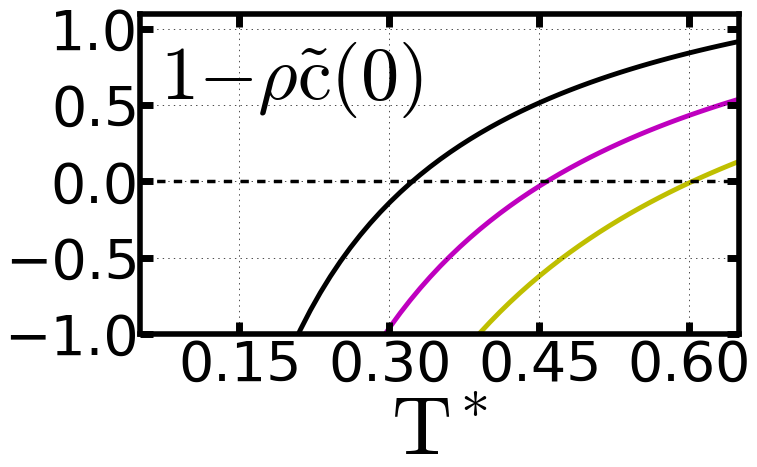}
\caption{\label{fig:chi_temp}(Color online)  Numerical solutions to Eq.\ref{eq:chi} as function of $T^*$ for density $\rho^*=0.3$ and 
charge separations $\delta=0.1,0.21,0.3\sigma$ colored in yellow, purple and black, respectively.}
\end{figure}
To further justify these interpretations, particularly the emergence of fluid phases, 
we performed a stability analysis of the homogenous isotropic high temperature state
based on mean-field density functional theory (DFT).
Specifically, we consider the isothermal compressibility $\chi_T$.
Positive values of $\chi_T$ imply that the homogeneous (fluid) phase is stable, 
whereas negative values indicate that this phase is unstable.
Specifically, the instability arises
against long-wavelength density fluctuations, i.e. condensation.
According to Kirkwood-Buff theory~\cite{Kirkwood_Buff} one has
\begin{equation}
\chi_T^{-1}\propto 1-\rho\tilde c(k=0),
\label{eq:chi}
\end{equation}
where $\tilde{c}(0)$ is the Fourier transform of the direct correlation function (DCF) $c(\boldsymbol{r}_{12})$
in the limit of long-wavelengths ($k\rightarrow 0$).
We approximate the DCF for distances $r_{ij}>\sigma$ according to a mean field (MF) approximation, that is 
\begin{equation}
c_{MF}(\mathbf{r}_{12})=-(k_BT)^{-1}U(\mathbf{r}_{12}), \;\;\; r_{12}>\sigma,
\label{eq:DCFU}
\end{equation}
and use the Percus-Yevick DCF $c_{HS}(r_{12})$ of a pure hard-sphere fluid~\cite{DCF} for $|\mathbf{r}_{12}|\leq\sigma$.
The full DCF is then given by
\begin{equation}
c(\mathbf{r}_{12})=c_{HS}(r_{12})+c_{MF}(\mathbf{r}_{12}).
\label{eq:DCF}
\end{equation}
In Fig.~\ref{fig:chi_temp} we present numerical results for the 
expression $1-\rho\tilde c(0)$ at $\rho^*=0.3$ as function of temperature.
At low $T^{*}$, all systems are characterized by {\em negative} values of  $1-\rho\tilde c(0)$. 
This indicates that the homogeneous isotropic phase is unstable,
consistent with the results of our simulations. Upon increasing $T^{*}$
the mean-field compressibility $\chi_T$ then becomes indeed positive for all 
charge separations considered. Specifically, for $\delta=0.3\sigma$ the change of sign 
(related to a "spinodal point") occurs at $T_{f,sq}^*=0.325$ and for $\delta=0.1\sigma$ at the much 
higher temperature $T_{f,hex}^*=0.6$. 
These values are in surprisingly good agreement with our estimates for the "fluidization" temperatures
based on the order parameter plots.

The case $\delta=0.21\sigma$ is again different.
Here we find [see Fig.~\ref{fig:7}(b)] that, starting from low temperatures inside the DLCA regime, 
the mean coordination number monotonically decreases. 
However, this does not indicate "fluidization" but rather a {\em gradual} transition from a state with 
dominant hexagonal order towards a mixed state comprised
of coexisting clusters with local hexagonal and square-like order.
Indeed, [see Fig.~\ref{fig:7}(c)], the orientational order parameters $\Phi_4$ and $\Phi_6$ reveal 
that the fraction of particles bound in square clusters increases with $T^*$ and finally overtakes the 
fraction of particles involved in hexagonal clusters at $T^* \approx 0.35$. 
Corresponding snapshots of simulation results are shown in Fig.~\ref{fig:8}. 
At all temperatures considered one observes separated clusters. 
With increasing temperature their shape becomes more regular, while the local rectangular order 
becomes more pronounced.
Finally, at $T^*=0.45$ the fractal dimension $D_f$ and the 
square order parameter $\Phi_4$ reach their maximum values, suggesting
a "fluidization" similar to the behavior observed at other values of $\delta$.
Interestingly, our stability analysis [see Eq.~(\ref{eq:chi})] 
indicates an instability at the same temperature $T_f^*=0.45$. 
With this surprisingly accurate agreement between theory and simulation, we conclude that in 
the transition regime ($\delta=0.21\sigma$), increasing thermal fluctuations first push the system 
from a dominantly hexagonal state into a rectangular one, which then enters a 
metastable fluid phase after passing the "spinodal point".

\subsection{Spotlight on higher densities}
\label{Spotlight on higher densities}

\begin{figure}[]
\centering
\includegraphics[width=0.5\textwidth]{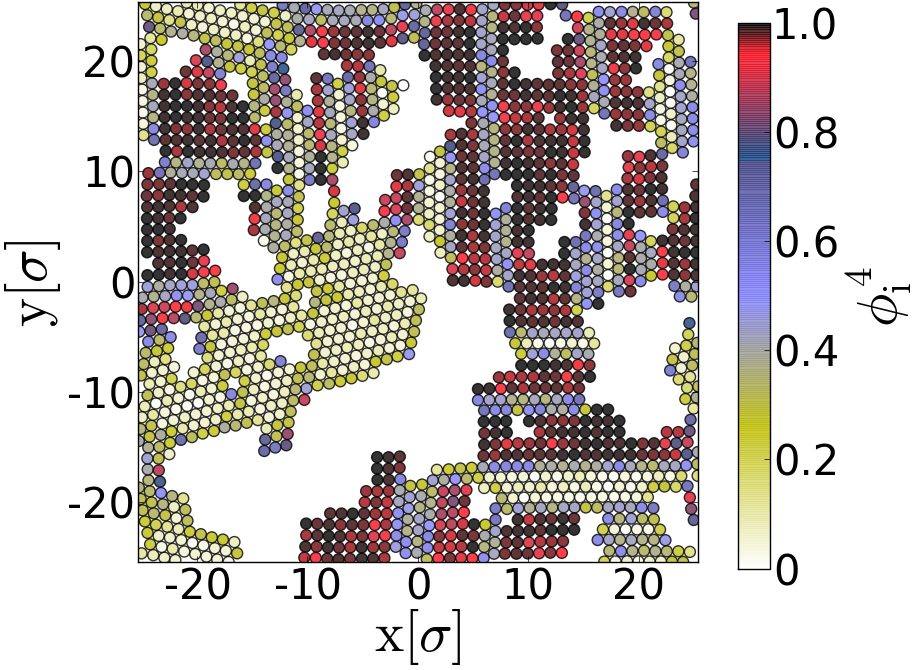}
\caption{\label{fig:9}(Color online)  System at $T^*=0.05$, density $\rho^*=0.7$ and $\delta=0.21\sigma$. 
The colorcode gives the orientational bond-order parameter $\phi_i^4$ of each particle $i$.}
\end{figure}

In this section we revisit the system behavior at the low temperature $T^*=0.05$, but consider different densities in the range $\rho^*\leq 0.7$. 
Whereas low-density systems at $T^{*}=0.05$ display DLCA as discussed in sections~\ref{Results}~A-C, this aggregation mechanism is expected to disappear at higher densities: here,
the particles are just unable to diffuse sufficiently freely. 
Rather, the particles will very frequently collide and then immediately form rigid bonds.
A typical structure at the highest density considered, $\rho^*=0.7$, and separation parameter $\delta=0.21\sigma$ is shown in Fig.~\ref{fig:9}. 
Clearly, the system is percolated, that is, the particles form a single, system-spanning cluster. Interestingly, this cluster is composed of extended regions 
characterized by either square-like order or hexagonal order. We note that, at $\delta=0.21\sigma$, 
simultaneous appearance of clusters with both types of order also occurs at 
low densities and higher temperatures (see section~\ref{Beyond DLCA - Higher Temperatures}). 
However, at the high density considered here the regions of each type are larger and 
the particle arrangements are much more regular (i.e., there are less defects).

\begin{figure}[]
\centering
\includegraphics[width=0.49\textwidth]{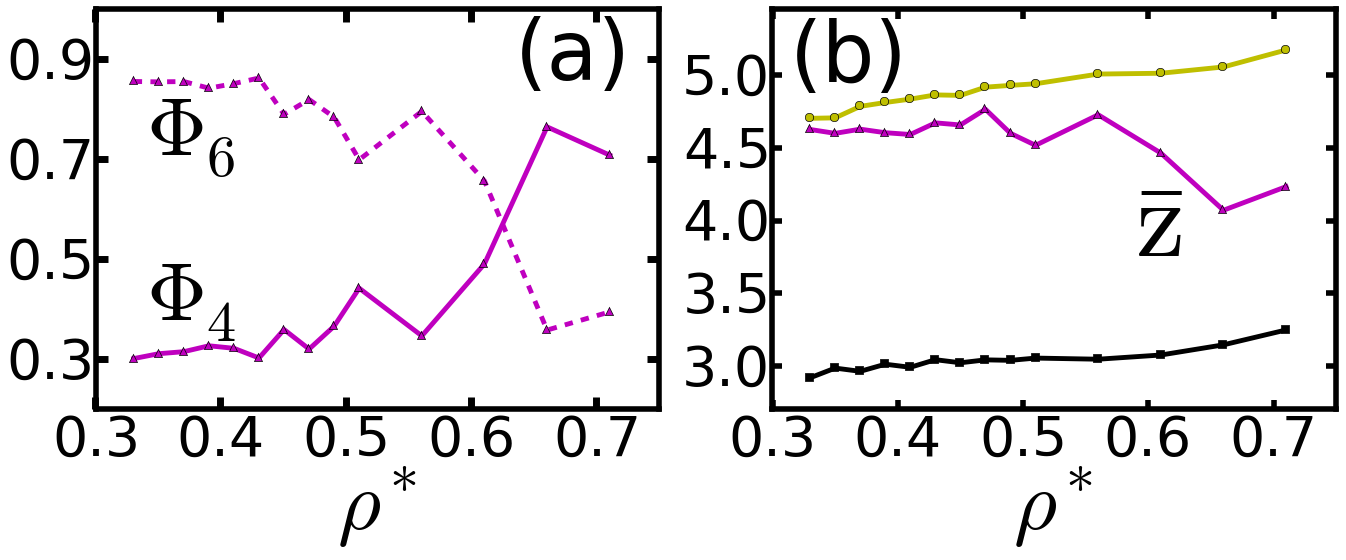}
\caption{\label{fig:10}(Color online) % 
(a) Orientational bond order parameter $\Phi_{4}$ and $\Phi_6$ as function of density $\rho^*$ for $\delta=0.21\sigma$ at $T^*=0.05$.  
(b) Mean coordination number $\bar{z}$ as function of $\rho^*$ at $T^*=0.05$ for different $\delta=0.1\sigma,0.21\sigma$ and $0.3\sigma$ colored in yellow, 
purple and black, respectively.
}
\end{figure}

To better understand the impact of the density on the cluster structures we plot in Fig.~\ref{fig:10}(a) the orientational bond
order parameters $\Phi_4$ and $\Phi_6$ as functions of $\rho^{*}$ for $\delta=0.21\sigma$ 
(at $\delta=0.1\sigma$ and $\delta=0.3\sigma$ the order parameters are essentially independent of the density). 
From Fig.~\ref{fig:10}(a) it is seen that 
the amount of rectangular (hexagonal) order sharply increases (decreases) at a density of $\rho^{*}\approx 0.45$. 
This is a surprising result as one would expect that, upon compressing the system, close-packed, hexagonal structures rather become more likely. 
However, at the low temperature considered here, structural reorganization is strongly hindered. 

We also note that  {\em all} of the systems investigated at densities  $\rho^*>0.45$ turned out to be percolated
(suggesting that the value $\rho^*=0.45$ is indeed related to the percolation transition).
It thus seems that the percolation tends to stabilize the initially formed square-lattice symmetry, as the subsequent reorganization is hindered
by the lack of mobility. In effect, we are faced with quenched states that could not densify within the time domain studied.
This interpretation is also consistent with the decrease of the mean coordination number once the system is
percolated ($\rho^*>0.45$) as shown in Fig.~\ref{fig:10}(b).

\section{Conclusions}
\label{Conclusions}
In this work we propose a new model for field-directed aggregation of colloidal particles with anisotropic interactions induced by external fields. The model
was inspired by recent experimental work~\cite{kretzschmar2013-ironcap,orlinBhuvnesh,Bhuvnesh_nets,Velev2004-fielddriectedassembly,gangwal:nets,Biswal2013-magneticfractal} 
on novel colloidal particles in which external fields can induce two, essentially decoupled, dipoles.

The formation of particle networks with multiple percolation directions 
can find application in a range of new materials with anisotropic electrical 
and thermal conduction, magnetic or electric polarizability or unusual rheological 
properties. The aggregated clusters can be dispersed in liquid, while the percolated 
networks can be embedded in a polymer or gel medium~\cite{tracy2013_percolated_in_polymer}. 
The key to the fabrication of 
such novel classes of materials containing particle clusters and networks is the 
control of the process parameters to obtain the desired 
interconnectivity, density and structure.

\balance
Against this background, the focus of our theoretical study was to understand the formation of {\it transient}, 
aggregated structures appearing at low-temperatures. Performing large-scale BD simulations
we have found that, depending on the distribution of the field-induced attractive "sites" in the particles, different aggregation mechanisms arise. 
These have been analyzed via appropriate structural order parameters, bond time-correlation functions as well as by the fractal dimension.
Our BD results demonstrate that by varying the charge separation parameter, that is, the distribution of attractive sites, the systems transform
from DLCA (essentially rigid bonds) towards sDLCA (slippery bonds).
Moreover, we show that
the change of aggregation behavior is accompanied by significant changes of the local cluster structure.

Indeed, the cluster structure can be easily manipulated by 
exploiting the interplay between temperature, density and model parameter $\delta$. This allows formation of unexpected structures
e.g., pronounced rectangular packing instead of closed packed hexagonal structures by increasing density. 
This unusual behavior appears to be dictated by the inability 
of the originally formed lattices with square symmetry to re-arrange into more 
dense hexagonal lattices. It has potentially important consequences for colloidal 
assembly, as is points out the ability to use mutidirectional field-driven assembly 
for the making of lower-density, yet highly interconnected, phases. 

Future research should focus on a more detailed investigation of the interplay between the aggregation mechanisms observed here (anisotropic and slippery DLCA),
and the {\it equilibrium} phase behavior, particularly the location of a condensation transition and of percolation at higher densities. 
This includes investigation of the influence of entropy which we did not discuss but is expected to strongly influence the aggregation behavior~\cite{Granick-entropyopenlattice}.

In conclusion, our study makes an important contribution to current research on structural and dynamical phenomena accompanying self-assembly
of complex colloids~\cite{Sciortino2008-irreversible,Sciortino2010-networkfluid}.
In particular, we have introduced a generic model describing colloids in multidirectional fields yielding tunable multipolar interactions.
Our study thus provides a microscopic understanding of aggregation processes in such systems.

\section*{Acknowledgements}
We gratefully acknowledge stimulating discussions with Bhuvnesh Bharti (NCSU). 
This work was supported by the NSF's Research Triangle MRSEC, DMR-1121107 and by The German Research Foundation via the International Research Training Group (IRTG) 1524
'Self-Assembled Soft Matter Nano-Structures at Interfaces'.

\end{document}